\documentclass[letterpaper]{article} % DO NOT CHANGE THIS
\usepackage{aaai24}  % DO NOT CHANGE THIS
\usepackage{times}  % DO NOT CHANGE THIS
\usepackage{helvet}  % DO NOT CHANGE THIS
\usepackage{courier}  % DO NOT CHANGE THIS
\usepackage[hyphens]{url}  % DO NOT CHANGE THIS
\usepackage{graphicx} % DO NOT CHANGE THIS
\usepackage{natbib}  % DO NOT CHANGE THIS AND DO NOT ADD ANY OPTIONS TO IT
\usepackage{caption} % DO NOT CHANGE THIS AND DO NOT ADD ANY OPTIONS TO IT
\usepackage{bm}
\usepackage{amssymb}
\usepackage{enumitem}
\usepackage{amsmath}
\usepackage{subfigure}
\usepackage{multirow}
\usepackage{appendix}
\usepackage{enumitem}
\usepackage[ruled, vlined, boxed, linesnumbered]{algorithm2e}
\usepackage{bibentry}
\usepackage{newfloat}
\usepackage{listings}
\DeclareCaptionStyle{ruled}{labelfont=normalfont,labelsep=colon,strut=off} % DO NOT CHANGE THIS
\title{Adversarial Socialbots Modeling Based on Structural Information Principles}
\author{
    Xianghua Zeng\textsuperscript{\rm 1},
    Hao Peng\textsuperscript{\rm 1},
    Angsheng Li\textsuperscript{\rm 1,2}
}
\affiliations{
    \textsuperscript{\rm 1} State Key Laboratory of Software Development Environment, Beihang University, Beijing, China\\
    \textsuperscript{\rm 2} Zhongguancun Laboratory, Beijing, China\\
    \{zengxianghua, penghao, angsheng\}@buaa.edu.cn, 
liangsheng@gmail.zgclab.edu.cn
}
\usepackage{bibentry}
\begin{document}

\maketitle

\begin{abstract}
The importance of effective detection is underscored by the fact that socialbots imitate human behavior to propagate misinformation, leading to an ongoing competition between socialbots and detectors.
Despite the rapid advancement of reactive detectors, the exploration of adversarial socialbot modeling remains incomplete, significantly hindering the development of proactive detectors.
To address this issue, we propose a mathematical \textbf{S}tructural \textbf{I}nformation principles-based \textbf{A}dversarial \textbf{S}ocialbots \textbf{M}odeling framework, namely \textbf{SIASM}, to enable more accurate and effective modeling of adversarial behaviors.
First, a heterogeneous graph is presented to integrate various users and rich activities in the original social network and measure its dynamic uncertainty as structural entropy. 
By minimizing the high-dimensional structural entropy, a hierarchical community structure of the social network is generated and referred to as the optimal encoding tree.
Secondly, a novel method is designed to quantify influence by utilizing the assigned structural entropy, which helps reduce the computational cost of SIASM by filtering out uninfluential users.
Besides, a new conditional structural entropy is defined between the socialbot and other users to guide the follower selection for network influence maximization.
Extensive and comparative experiments on both homogeneous and heterogeneous social networks demonstrate that, compared with state-of-the-art baselines, the proposed SIASM framework yields substantial performance improvements in terms of network influence (up to $16.32\%$) and sustainable stealthiness (up to $16.29\%$) when evaluated against a robust detector with $90\%$ accuracy.
\end{abstract}

\section{Introduction} \label{section: introduction}
Socialbots, automated user accounts partly controlled by software, have become indispensable in manipulating public opinion on social media \cite{ferrara2016rise, subrahmanian2016darpa}.
However, socialbots often receive criticism for disseminating false or unreliable information, causing confusion among online communities regarding crucial issues \cite{das2016manipulation, deb2019perils, aral2019protecting}.
Differentiating socialbots from regular user accounts is challenging due to their diverse and dynamic behaviors that resemble real-life users' behaviors, including forming appropriate followers and engaging in interactions \cite{cresci2020decade,arin2023deep}.

Various methods have been proposed to detect socialbots and prevent their negative impact, including supervised and unsupervised machine learning approaches \cite{chavoshi2016debot, varol2017online, yang2023fedack}.
Researchers have developed a framework that relies on minimal account metadata to improve scalability and generalization for adequate detection \cite{yang2020scalable}.
Nevertheless, these methods are passive as they wait for socialbot evasion before developing appropriate detection measures \cite{cresci2021coming}.
Instead of relying on these reactive approaches, researchers explore proactive detection methods using the multi-agent hierarchical reinforcement learning (HRL) mechanism \cite{le2022socialbots}, which simulates and understands the adversarial behaviors of socialbots.
Within this HRL mechanism, two adversarial objectives of socialbots are formulated: to survive under robust detectors by determining activity types and to maximize network influence by selecting good followers.
However, implementing the HRL mechanism poses practical challenges.
On the one hand, it is insufficient for the upper-level agent to dynamically determine activity types solely by learning from scratch without considering the entire social network structure.
On the other hand, the lower-level agent maintains the local features of all users to select followers, which leads to computational inefficiency.

In this paper, we propose a novel mathematical \textbf{S}tructural \textbf{I}nformation principles-based \textbf{A}dversarial \textbf{S}ocialbots \textbf{M}odeling framework, namely \textbf{SIASM}, to address the challenges above and further develop proactive detection.
Firstly, we transform the diverse user nodes and their multi-relational interconnections in the original social network into a unified structure, specifically, a heterogeneous graph.
We calculate the structural entropy of this graph to quantify its dynamic uncertainty.
Secondly, we minimize the high-dimensional structural entropy to generate an optimal encoding tree representing a hierarchical community structure of social users.
Each node in this tree corresponds to a user community, where users of the same community engage in frequent interactions.
To enhance the computational efficiency of SIASM, thirdly, we present a new method for quantifying the network influence of each community.
This method utilizes the assigned structural entropy of the corresponding tree node to filter out trivial communities with low influence.
Fourthly, we define a conditional structural entropy measure between the socialbot node and each user node, which guides the selection of appropriate followers to maximize network influence.
Extensive synthetic and real-life social network experiments are conducted to evaluate our proposed framework’s network influence and sustainable stealthiness.
Comparative results and analyses demonstrate its performance advantages over state-of-the-art baselines.
Furthermore, all source codes and experimental results are available at an anonymous link\footnote{\url{https://github.com/SELGroup/SIASM}}.

In summary, the contributions of our work can be summarized as follows:

$\bullet$ An innovative structural information principles-based framework, called SIASM, is proposed to tackle the challenges of insufficiency and inefficiency in adversarial socialbot modeling.

$\bullet$ A novel method for quantifying network influence using the assigned structural entropy of each user community is presented to effectively filter out uninfluential users and reduce the computational complexity of SIASM.

$\bullet$ A new conditional structural entropy between the socialbot and each user node is defined to guide the follower selection and maximize the network influence of socialbot.

$\bullet$ Our experiments on social networks demonstrate that SIASM achieves significant improvements of up to $116.64(16.32\%)$ and $13.21(16.29)\%$ in network influence and sustainable stealthiness compared to SOTA baselines.

\section{Preliminaries} \label{section: preliminaries}
\subsection{Social Network Environment}
In this work, we model the social network environment by referring to the problem formulations from the ACORN method \cite{le2022socialbots}, which includes network representation, diffusion model, and socialbot.

\textbf{Network Representation.}
A social network is modeled as a directed multi-relational graph $G_m=(V,\{\mathcal{E}_a\}_{a \in \mathcal{A}})$, where $V$ is the set of vertices\footnote{Vertices are defined in the graph, and nodes are in the tree.} representing social users, $\{\mathcal{E}_a\}_{a \in \mathcal{A}}$ is the set of edges representing various social activities $\mathcal{A}$.
When a social activity $a \in \mathcal{A}$ occurs between vertices $v_i$ and $v_j$, we construct a directed edge $e_{i, j}^{a} \in \mathcal{E}_{a}$, which signifies the transmission of news from $v_i$ to $v_j$, thereby $v_i$ influencing $v_j$.

\textbf{Influence Diffusion Model.}
Similar to previous studies on social networks \cite{jendoubi2017two, li2017survey}, we adopt the Independence Cascade Model (ICM) \cite{goldenberg2001talk} to represent the propagation of real-life news or influence in the graph $G_m$.
In the ICM, a specific set of vertices known as initial followers $F$ are active, while the rest remain inactive.
At each discrete timestep, each active vertex $v \in F$ has an equal probability $p$ of activating its inactive neighbors $\mathcal{N}(v)$.
Once no additional vertices are left to activate, the propagation process concludes \cite{kamarthi2020influence, li2021claim}.
We use the notation $\sigma(G_m, p)$ to denote the number of vertices a piece of news can reach from $F$ through the ICM model.

\textbf{Socialbots.} In the graph $G_m$, a socialbot $b \in B$ refers to a vertex that imitates real-life behaviors with the aim of spreading misinformation or low-credible content.
Its adversarial objectives mainly consist of two parts: 1) optimizing network influence by selecting good user nodes as followers $F \subset V$ over time, and 2) evading detection and removal by robust bot detectors.

\subsection{Markov Decision Process}
To simultaneously optimize the above objectives, we model the adversarial behaviors of socialbots as a Markov Decision Process (MDP) \cite{bellman1957markovian}, denoted by $\mathcal{M}=(\mathcal{S}, \mathcal{A}, \mathcal{R}, \mathcal{P}, \gamma)$.
This process consists of a state space $\mathcal{S}$, an action space $\mathcal{A}$, a transition function $\mathcal{P}$, a reward function $\mathcal{R}$, and a discount factor $\gamma \in [0, 1]$.
At each timestep, the agent receives an environmental state $s \in \mathcal{S}$ and chooses an action $a \in \mathcal{A}$ based on its policy $\pi(s, a)$, resulting in a new state $s^\prime \sim \mathcal{P}(s, a)$ and a reward $r \sim \mathcal{R}(s, a)$.

\subsection{Structrual Information Principles}
A partition of the vertices set $V$ in a homogeneous graph $G=(V, E)$ is defined as $P=\{P_0, P_1,\dots\}$, where each $P_i$ is a community that serves as a cluster of vertices. 
And these communities can be further divided into sub-communities in a hierarchical manner.
Unlike traditional information entropy used in communication systems \cite{shannon1948mathematical}, Li and Pan \cite{li2016structural} first proposed structural entropy to quantify the dynamic certainty embedded in complex networks under such hierarchical partitions.
In this work, the hierarchical partitions are represented by a tree structure known as the encoding tree.

\textbf{Encoding Tree.}
Similar to the previous study \cite{zeng2023effective}, the encoding tree $T$ of graph $G$ is formally defined as a rooted tree with the following properties:
1) For the root node $\lambda$, the set of vertices corresponding to $\lambda$ is denoted as $V_\lambda=V$.
2) For each leaf node $\nu$, the set consists of a single vertex $v \in V$, represented as $V_\nu=\{v\}$.
3) For each non-root and non-leaf node $\alpha$, there exists a subset of vertices $V_\alpha$ corresponding with $\alpha$, and its parent node is denoted as $\alpha^{-}$. 
4) For each non-leaf node $\alpha$, we assume the number of its children as $L_\alpha$ and its $i$-th child as $\alpha^{\langle i \rangle}$, respectively.
5) For each non-leaf node $\alpha$, all subsets of vertices $V_{\alpha^{\langle i \rangle}}$ are disjointed and set $V_\alpha=\bigcup_{i=1}^{L_\alpha}V_{\alpha^{\langle i \rangle}}$.

\textbf{One-dimensional Structural Entropy.}
Without any hierarchical partitioning structure, the dynamic certainty of graph $G$ is measured as the one-dimensional structural entropy and defined as follows:
\begin{equation}\label{1d_se}
    H^{1}(G)=-\sum_{v \in V} \frac{d_v}{vol(G)} \cdot \log_2 \frac{d_v}{vol(G)}\text{,}
\end{equation}
where $d_{v}$ is the degree of vertex $v$ and $vol(G)=\sum_{v \in V} d_{v}$ is the volume of $G$.

\textbf{High-dimensional Structural Entropy.}
An encoding tree $T$ can significantly reduce the dynamic uncertainty of graph $G$, and the high-dimensional structural entropy measures the remaining uncertainty embedded in $G$.
For each non-root tree node $\alpha \in T$, its assigned structural entropy is defined as follows:
\begin{equation}\label{kd_se_node}
    H^{T}(G;\alpha)=-\frac{g_{\alpha}}{vol(G)} \log _{2} \frac{\mathcal{V}_{\alpha}}{\mathcal{V}_{\alpha^{-}}}\text{,}
\end{equation}
where $\mathcal{V}_\alpha$ is the volume of $V_\alpha$ and $g_\alpha$ is the sum of all edge weights connecting each vertex in $V_\alpha$ and each vertex outside $V_\alpha$.
The $K$-dimensional structural entropy is defined:
\begin{equation}\label{node_se}
    H^T(G)=\sum_{\alpha \in T, \alpha \neq \lambda}H^{T}(G;\alpha)\text{,}
\end{equation}
\begin{equation}\label{kd_se}
    H^{K}(G)=\min_{T}\left\{H^T(G)\right\}\text{,}
\end{equation}
where $T$ ranges over all encoding trees whose height are at most $K$, $K > 1$.

\section{The Proposed SIASM Framework} \label{section: ASMSI}
\begin{figure}[t]
    \centering
    \includegraphics[width=1\columnwidth]{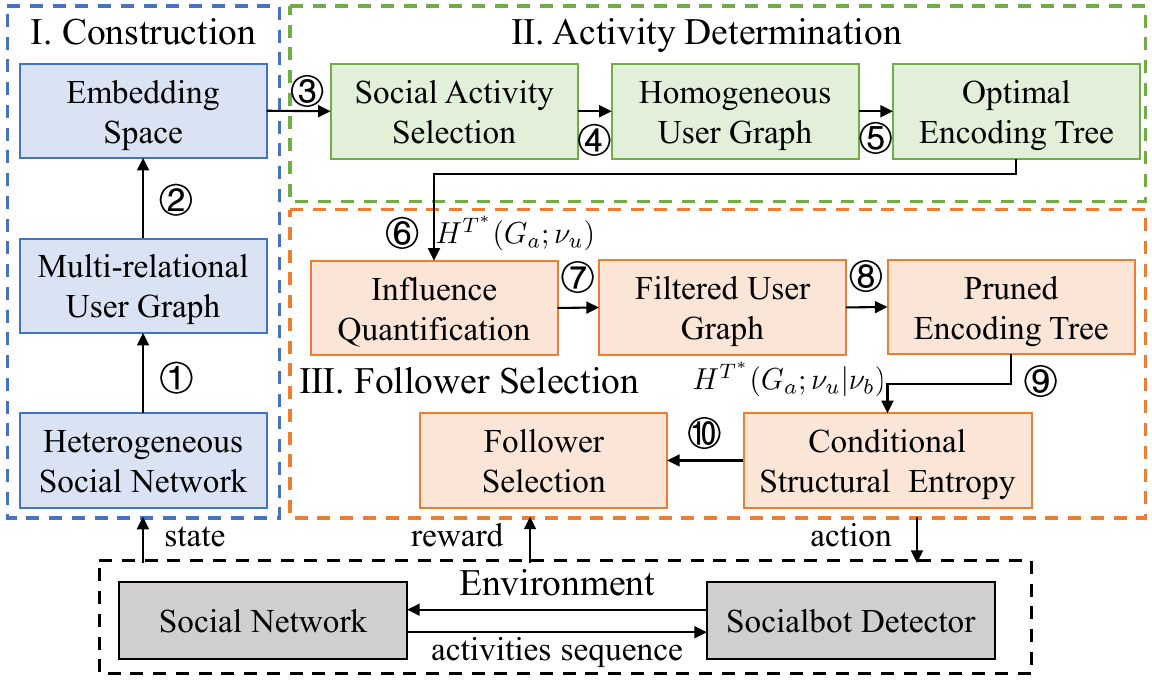}
    \caption{The proposed adversarial socialbot modeling framework: SIASM.}\label{fig:SIASM Framework}
\end{figure}

\begin{figure*}[t]
    \centering
    \includegraphics[width=1\textwidth]{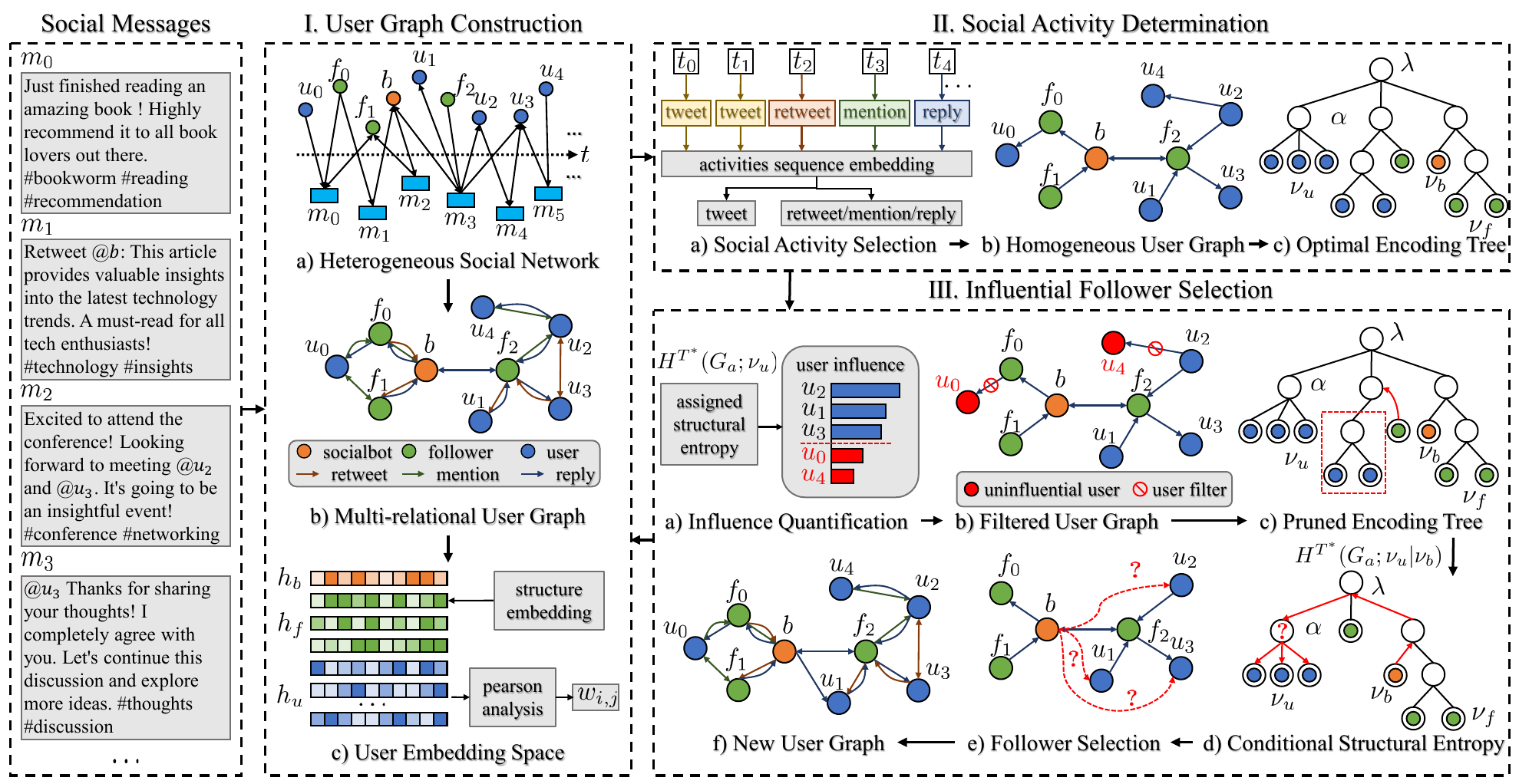}
    \caption{The detailed designs of SIASM framework
    .}\label{fig:SIASM}
\end{figure*}

In this work, we adopt the structural information principles to dynamically model the adversarial behaviors of socialbots, as shown in Fig. \ref{fig:SIASM Framework}.
The SIASM receives the environmental state of the social network, generates a social activity to update the network structure, and obtains reward information for further optimization.
Specifically, the architecture of SIASM encompasses three stages: graph construction, activity determination, and follower selection. 
In the graph construction stage, we transform the original heterogeneous social network into a multi-relational user graph and encode its network structure in an embedding space. 
In the activity determination stage, we select a social activity to simplify the multi-relational graph into a homogeneous graph and minimize its structural entropy to generate the optimal encoding tree. 
In the follower selection stage, we quantify the network influence of each user, remove trivial user vertices and tree nodes with low influence from the user graph and encoding tree, and measure conditional structural entropy to guide the follower selection.
Fig. \ref{fig:SIASM} showcases the detailed design of the proposed SIASM framework.

\subsection{User Graph Construction}
Originating from the history of social messages, we extract user elements and various social activities, including $tweet$, $retweet$, $mention$, and $reply$, to construct the heterogeneous social network shown in Fig. \ref{fig:SIASM}.
To preserve heterogeneous information across different elements, we convert the social graph into a multi-relation user graph, denoted as $G_m=(V, \{\mathcal{E}_a\}_{a \in \mathcal{A}})$ in Fig. \ref{fig:SIASM}.
The vertices $V$ in this multi-relational graph represent social users, including socialbots $B \subset V$ and followers $F \subset V$.
These collections are equipped with pre-trained features of word embedding and timestamp encoding, enhancing their semantic representations and temporal information.
When users share the same message with different social activities $\mathcal{A}$, edges representing various user relations are established in $G_m$.
As the levels of impurities differ across these relations within the multi-relational graph and collectively impact the embedding results, we adopt the R-GCN network \cite{schlichtkrull2018modeling} to encode the structure of $G_m$.
For each user vertex $v \in V$, its pre-trained features and the multi-relational structural information are integrated to encode it into a $d$-dimensional embedding $h_v$ in Fig. \ref{fig:SIASM}, effectively enhancing its representation.

\subsection{Social Activity Determination} \label{section: SAD}
In this stage, SIASM introduces an RL agent that dynamically selects a type of social activity to simplify the multi-relational graph, transforming it into a homogeneous user graph and minimizing its structural entropy.
This process generates a hierarchical community structure of social users for subsequent influential follower selection.

At each timestep, the RL agent encodes its activities history as a fixed-dimensional vector and utilizes the learned representation to determine which social activity to perform, aiming for sustainable stealthiness, as depicted in step \uppercase\expandafter{\romannumeral2}. a in Fig. \ref{fig:SIASM}.
If the $tweet$ is chosen, the influential follower selection stage is skipped within the SIASM framework for that particular timestep.
Based on the chosen activity $a$, we simplify the multi-relational graph $G_m$ into a homogeneous user graph $G_a=(V,\mathcal{E}_a)$ in Fig. \ref{fig:SIASM} by eliminating directed edges representing other types of social activities.
To calculate the weight $w_{i, j}\in\left[-1, 1\right]$ for each directed edge $e^{a}_{i,j}=(v_i,v_j)$, we employ Spearman Correlation Analysis on their representations $h_{v_i}$ and $h_{v_j}$ as follows:
\begin{equation}\label{pcc}
    w_{i j}=1-\frac{6 *\left\|k\left(h_{v_i}\right)-k\left(h_{v_j}\right)\right\|_2^2}{d *\left(d^2-1\right)}\text{,}
\end{equation}
where $k(h_v)$ is the rank of the $d$-dimensional vertex representation $h_{v}$.

\begin{algorithm}[t]
    \SetAlgoVlined
    \KwIn{The one-layer initial encoding tree $T$, $K \in \mathbb{Z}^{+}$}
    \KwOut{The $K$-layer optimal encoding tree $T^*$}
    $h_{T} \gets$ the height of $T$\\
    \While{$h_T < K$} {
        $i^* \gets \arg \max_{i}\{\overline{R_{se}}(T;U_i)\}$\\
        \If{$\overline{R_{se}}(T;U_{i^*})=0$} {
            break
        }
        \For{$\alpha \in U_{i^*}$} {
            \textit{stretch} $T_\alpha$\\
            \textit{compress} $T_\alpha$\\
            $h_T \gets h_T + 1$
        }
        \For{$i=i^* + 1, ..., h_T$} {
            update $U_{i}$
        }
    }
    $T^{*} \gets T$\\
    \Return{$T^{*}$}
\caption{The Optimization Algorithm}
\label{algorithm:optimization}
\end{algorithm}

To assess the impact of a specific activity $a$ on the social network, we model message propagation as a random walk between users in the homogeneous graph $G_a$ and incorporate structural entropy to quantify the inherent dynamic uncertainty.
We generate an optimal encoding tree as a hierarchical partitioning structure of user communities by minimizing its high-dimensional structural entropy.
Specifically, we initialize a one-layer encoding tree $T$ for the homogeneous graph $G_a$ as follows: (1) For the entire set of vertices $V$, we generate a root node $\lambda$ and set $V_\lambda=V$; (2) For each vertex $v \in V$, we generate a leaf node $\nu$ and set $V_\nu=\{v\}$; (3) For each leaf node $\nu$, we assign its father as the root node $\lambda$, denoted as $\nu^{-}=\lambda$.
Additionally, we incorporate two operators, \textit{stretch} and \textit{compress}, from the HCSE algorithm \cite{pan2021information} to optimize the one-layer encoding tree.
In our work, we denote the average reduction of structural entropy resulting from one round of \textit{stretch} and \textit{compress} operations on all tree nodes $U_i$ in layer $i$ as $\overline{R_{se}}(T; U_i)$.
We generate the optimal encoding tree $T^{*}$ with $K$ layers by iteratively and greedily selecting tree nodes to execute the above operations.
The optimization process is summarized in algorithm \ref{algorithm:optimization}.
In each iteration, we examine all sets of nodes at the same layer to identify the set $U_{i^*}$ that maximizes the average reduction of structural entropy $\overline{R_{se}}$ (line 3 in algorithm \ref{algorithm:optimization}) and then execute \textit{stretch} and \textit{compress} operations for each node in $U_{i^*}$ (lines 7 and 8 in algorithm \ref{algorithm:optimization}).
We continue these operations until either the tree height $h_T$ reaches $K$ (line 2 in algorithm \ref{algorithm:optimization}) or no nodes are satisfying $\overline{R_{se}}(T;U_{i^*})>0$ (line 4 in algorithm \ref{algorithm:optimization}).
We terminate the iteration and output $T$ as the optimal encoding tree $T^*$.

The encoding tree $T^*$ in Fig. \ref{fig:SIASM} illustrates the hierarchical community structure of social users, where each tree node represents a user community, and its height signifies the community's position within the hierarchy.
Each leaf node $\nu$ corresponds to a singleton consisting of a single user vertex $v$, with $V_\nu=\{v\}$.
Each non-leaf node $\alpha$ corresponds to a new community $V_\alpha$, which consists of the communities of its children, $V_\alpha=\bigcup_{i=1}^{L_\alpha}V_{\alpha^{\langle i \rangle}}$.
The root node $\lambda$ corresponds to the entire set of social users, with $V_\lambda=V$.

\subsection{Influential Follower Selection}
After determining the activity type in the previous stage, the SIASM quantifies the network influence of each user community and calculates the conditional structural entropy between the socialbot and each leaf node to guide follower selection for maximizing its social influence.

For each tree node $\alpha$ in the optimal encoding tree $T^*$, we calculate its assigned structural entropy $H^{T^*}(G_a;\alpha)$ using Eq. \ref{node_se}.
Intuitively, a higher value of $H^{T^*}(G_a;\alpha)$ indicates a greater likelihood of a specific type of social activity $a$ occurring between users within community $V_\alpha$ compared to other communities starting at the siblings of $\alpha$.
In step \uppercase\expandafter{\romannumeral3}. a of Fig. \ref{fig:SIASM}, we quantify the network influence $I_\alpha$ of user community $V_\alpha$ by summing up the likelihood of occurrence for each tree node $\beta$ on the path from the root $\lambda$ to the node $\alpha$ as follows:
\begin{equation}
    I_\alpha=\sum_{V_{\alpha} \subseteq V_\beta \subset V_\lambda} H^{T^*}\left(G_a; \beta\right)\text{.}
\end{equation}
To reduce the problem size of follower selection, we prune branches starting at nodes $\alpha$ with low network influence in $T^*$ and filter out all users in the community $V_\alpha$ from the graph $G_a$, as steps \uppercase\expandafter{\romannumeral3}. b and \uppercase\expandafter{\romannumeral3}. c in Fig. \ref{fig:SIASM}.
This strategy effectively decreases the number of potential followers to select from, guaranteeing the SIASM framework's efficiency.
In this work, we set the ratio of filtered user vertices and the height of pruned subtrees as $5\%$ and $1$, respectively.

We trace the path of each potential follower $u$ from $\nu_u$ to the root $\lambda$ and verify if the socialbot $b$ exists in the community $V_\delta$ at every tree node $\delta$ on this path. 
When we locate a node $\delta$ encompassing both $\nu$ and $b$ within its community $V_\delta$, we calculate the conditional structural entropy $H^{T^*}(G_a;\nu_u \mid \nu_b)$ as step \uppercase\expandafter{\romannumeral3}. d in Fig. \ref{fig:SIASM}:
\begin{equation}
    H^{T^*}(G_a;\nu_u \mid \nu_b)=\sum_{V_{\nu_u} \subseteq V_\alpha \subset V_\delta} H^{T^*}\left(G_a; \alpha\right)\text{.}
\end{equation}
We leverage conditional structural entropy to measure the uncertainty from the father node $\delta$ to the leaf node $\nu_{u}$, where $\alpha$ is any node on this path.
This entropy reflects the probability of a piece of news originating from the socialbot $b$ reaching the user $u$, and we use it as the initial selection probability for maximizing the socialbot's network influence.
To further refine this probability, we employ the actor-critic RL (PPO) algorithm \cite{schulman2017proximal}.
For an efficient optimization process, we utilize the state abstraction mechanism \cite{zeng2023hierarchical} that extracts essential features from all vertices' representation vectors and enables effective follower selection, as step \uppercase\expandafter{\romannumeral3}. e in Fig. \ref{fig:SIASM}.
Finally, the socialbot adds the user $u$ as a new follower and construct a directed edge of social activity $a$, thereby yielding an updated user graph in Fig. \ref{fig:SIASM}.

\subsection{Time Complexity of SIASM}
In this section, we analyze the time complexity of the SIASM framework, which encompasses user graph construction, social activity determination, and social follower selection stages, to assess its practicality.
The overall complexity of SIASM is $O\left(m+n+n \cdot n_a+m \cdot  \log ^{2} n\right)$, where $n$ represents the number of users, $m$ the number of messages, and $n_a$ the number of social activities.
Specifically, in the graph construction stage, constructing the social network or multi-relational user graph takes $O\left(m+n\right)$ time complexity, and embedding user representations takes $O\left(m+n \cdot n_a\right)$ time complexity. 
In the activity determination stage, simplifying the multi-relational graph incurs a complexity of $O\left(n_a\right)$, and optimizing the encoding tree leads to a complexity of $O\left(m \cdot \log^2n\right)$.
During the follower selection stage, the SIASM requires $O\left(n\right)$ time complexity to quantify influence, filter user communities, and select appropriate followers.

\section{Experiments and Analysis} \label{section: experiment}
In this section, we present empirical and comparative experiments on homogeneous and heterogeneous social networks to validate the superiority of the SIASM framework. 
And we provide all experimental results, including their corresponding average values and standard deviations.
Each experiment is conducted with five random seeds to ensure unbiased evaluations and avoid discrepancies.

\subsection{Experimental Setup}
\subsubsection{Datasets.}
To analyze homogeneous datasets, we collect the top 100 trending articles about the US presidential election and COVID-19 pandemic topics from Twitter and study their propagation networks, which include 1500 social users.
For heterogeneous network analysis, we use the latest Higgs Twitter Dataset \cite{de2013anatomy}, which includes directed multi-relational interactions.
These networks comprise multiple star-shaped communities with limited connections, which share the same observations as previous research \cite{sadikov2011correcting, kamarthi2020influence}.
Like other works \cite{le2022socialbots}, we select $10\%$ of the real-life networks to construct synthetic stochastic networks as the training set and take the remaining $90\%$ of the collected networks as the testing set.

\subsubsection{Baselines.}
This work compares the SIASM with the state-of-the-art adversarial socialbots modeling method ACORN \cite{le2022socialbots}.
Additionally, we combine several classical heuristic approaches (CELF \cite{leskovec2007cost} and DEGREE \cite{chen2009efficient}) with the previously learned agent to create other baselines, known as ACRON-H and SIASM-H.

\subsection{Evaluations}
In this section, we evaluate various methods in both homogeneous and heterogeneous social networks using synthetic graphs to train models and real-life graphs to test their performances.
The resulting averages and deviations of episode rewards and lengths are summarized in Table \ref{fig:overall result}.
Our analysis shows that SIASM significantly improves episode rewards and lengths in both synthetic and real-life graphs.
Specifically, SIASM achieves up to $116.64 (16.32\%)$ and $13.21 (16.29\%)$ improvements in reward and length, demonstrating its advantages in network influence and sustainable stealthiness.
Furthermore, the following subsection separately shows a detailed analysis of experiments conducted on homogeneous or heterogeneous datasets.

\begin{table*}[t]
    \centering
    \resizebox{1\linewidth}{!}{
    \begin{tabular}{|c|c|c|c|c|c|c|}
    \hline
    \multirow{2}{*}{Homogeneous} & \multicolumn{2}{c|}{Synthetic Graph} & \multicolumn{2}{c|}{Real-life Graph} & \multicolumn{2}{c|}{Average Performance} \\ \cline{2-7}
         &  Episode Reward & Episode Length & Episode Reward & Episode Length & Episode Reward & Episode Length\\ \hline
         ACRON-H & - & - & $825.53 \pm 13.71 $ & $40.44 \pm 24.48$ & $\underline{825.53} \pm 13.71$ & $40.44 \pm 24.48$ \\ \hline
         SIASM-H & - & - & $820.84 \pm \underline{11.11}$ & $62.78 \pm 35.63$ & $820.84 \pm \underline{11.11}$ & $62.78 \pm 35.63$ \\ \hline
         ACRON   & $\underline{714.72} \pm \underline{59.49}$ & $\underline{97.12} \pm \underline{8.46}$ & $\underline{827.96} \pm 17.23$ & $\underline{67.89} \pm \underline{22.62}$ & $771.34 \pm 38.36$ & $\underline{82.51} \pm \underline{15.54}$ \\ \hline
         SIASM   & $\bm{831.36} \pm \bm{3.67}$ & $\bm{108.45} \pm \bm{2.01}$ & $\bm{831.02} \pm \bm{8.87}$ & $\bm{81.10} \pm \bm{20.16}$ & $\bm{831.19} \pm \bm{6.27}$ & $\bm{94.78} \pm \bm{11.09}$ \\ \hline
         Abs.($\%$) Avg. $\uparrow$ & $116.64(16.32\%)$ & $11.33(11.67\%)$ & $3.06(0.37\%)$ & $13.21(16.29\%)$ & $5.66(0.69\%)$ & $12.27(14.87\%)$ \\ \hline
    \multirow{2}{*}{Heterogeneous} & \multicolumn{2}{c|}{Synthetic Graph} & \multicolumn{2}{c|}{Real-life Graph} & \multicolumn{2}{c|}{Average Performance} \\ \cline{2-7}
         &  Episode Reward & Episode Length & Episode Reward & Episode Length & Episode Reward & Episode Length \\ \hline
         ACRON-H & - & - & $817.56 \pm 12.53$ & $74.36 \pm 10.42$ & $817.56 \pm 12.53$ & $74.36 \pm 10.42$ \\ \hline
         SIASM-H & - & - & $\underline{831.83} \pm \underline{7.66}$ & $\underline{75.39} \pm \underline{4.92}$ & $\underline{831.83} \pm \underline{7.66}$ & $75.39 \pm \underline{4.92}$ \\ \hline
         ACRON & $800.62 \pm 4.08$ & $\underline{108.71} \pm \underline{1.05}$ & $821.83 \pm 16.37$ & $71.73 \pm 11.36$ & $811.23 \pm 10.23$ & $\underline{90.22} \pm 6.21$ \\ \hline
         SIASM & $\bm{828.61} \pm \bm{2.16}$ & $\bm{110.61} \pm \bm{0.66}$ & $\bm{840.89} \pm \bm{7.26}$ & $\bm{80.0} \pm \bm{1.41}$ & $\bm{834.75} \pm \bm{4.71}$ & $\bm{95.31} \pm \bm{1.04}$ \\ \hline
         Abs.($\%$) Avg. $\uparrow$ & $27.99(3.50\%)$ & $1.9(1.75\%)$ & $9.06(1.09\%)$ & $4.61(6.11\%)$ & $2.92(0.35\%)$ & $5.09(5.64\%)$ \\ \hline
    \end{tabular}}
    \caption{Summary of overall experimental results in homogeneous and heterogeneous datasets: ``average value $\pm$ standard deviation" and ``improvements" ($\%$). Bold: the best performance in each graph, underline: the second performance.}
    \label{fig:overall result}
\end{table*}

\subsubsection{Homogeneous Datasets.}
\begin{figure}[t]
    \centering
    \includegraphics[width=1\columnwidth]{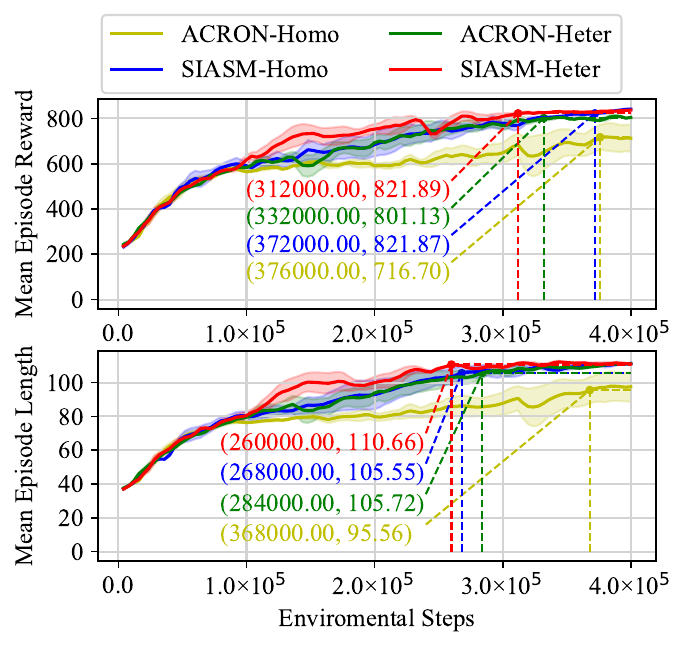}
    \caption{Learning curves of episode reward and episode length in homogeneous and heterogeneous social networks.}
    \label{fig:training}
\end{figure}

In each homogeneous social network, we do not consider different types of user activities such as \textit{tweet}, \textit{retweet}, \textit{mention}, and \textit{reply}.
Instead, we model them as a homogeneous user graph.
During the training process on synthetic graphs, we use a default propagation probability $(p)$ of $0.8$ and a maximal episode length $(T_{max})$ of $120$.
As depicted in Fig. \ref{fig:training}, our SIASM framework converges with fewer environmental steps ($312000$ and $260000$) and achieves better performances regarding network influence ($831.36$) and sustainable stealthiness ($108.45$) compared to other methods.
These advantages indicate the efficiency of SIASM in learning policies for follower selection and detection evasion, leading to desirable outcomes.

\begin{figure*}[t]
    \centering
    \includegraphics[width=1\textwidth]{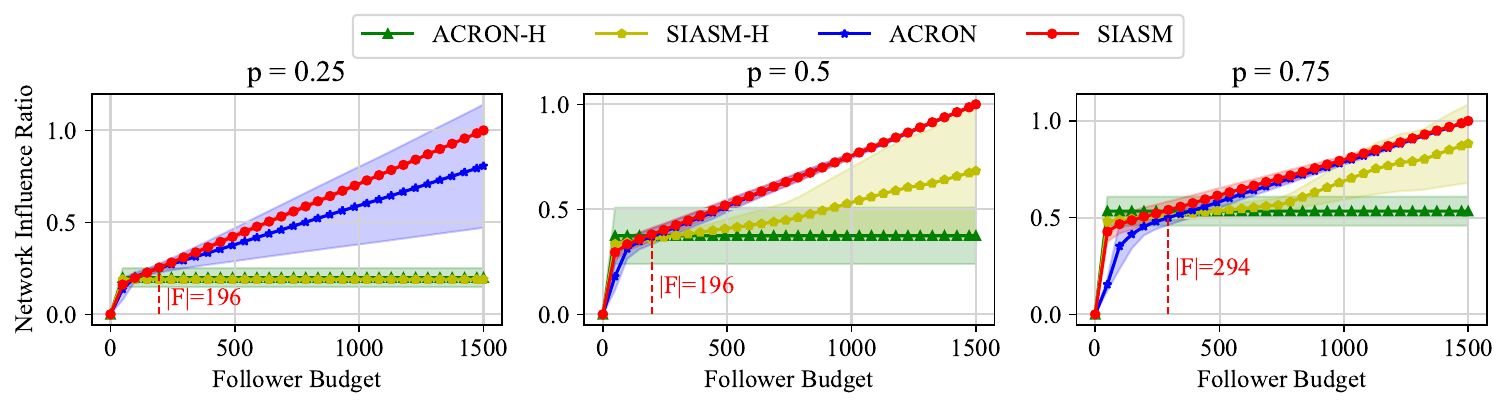}
    \caption{The testing process on homogeneous social networks}
    \label{fig:homo_test}
\end{figure*}

During the testing process using real-life graphs, we vary the $p$-value and measure the network influence ratio, which represents the ratio of users receiving target messages to all users across different follower budgets, as shown in Fig. \ref{fig:homo_test}.
Overall, SIASM consistently outperforms all baselines, particularly when the number of followers exceeds a threshold of $294$, $|F| > 294$.
It is evident that the SIASM stably achieves a more significant influence while interacting with fewer social users, regardless of propagation probabilities.
This superiority can be attributed to SIASM's ability to select suitable followers based on the global feature of structural entropy, thereby overcoming the limitations associated with baselines that rely on local features for node selection.

Furthermore, we conduct additional evaluations to assess the sustainable stealthiness of all compared methods. 
We measure their survival steps in real-life networks and summarize the comparative results in Table \ref{fig:ss}. 
The findings reveal that our SISAM demonstrates robust longevity, persisting for $2.4K$ timesteps with a network influence ratio of $0.99$ across various real-life scenarios, significantly outperforming other methods.

\begin{table}[t]
\centering
\resizebox{1\linewidth}{!}{
\begin{tabular}{|c|c|c|c|c|}
\hline
\multirow{2}{*}{Methods} & \multicolumn{2}{c|}{$p=0.25$} & \multicolumn{2}{c|}{$p=0.5$} \\ \cline{2-5} 
                  & \% $\uparrow$ & Steps $\uparrow$ & \% $\uparrow$ & Steps $\uparrow$ \\ \hline
ACRON-H           & $0.20 \pm 0.05$ & $1.2K \pm 1K$ & $0.37 \pm 0.13$ & $1.2K \pm 1K$ \\ \hline
SIASM-H           & $0.19 \pm \underline{0.03}$ & $1.8K \pm 996$ & $0.68 \pm 0.32$ & $1.1K \pm 1K$ \\ \hline
ACRON             & $\underline{0.81} \pm 0.34$ & $\underline{2.1K} \pm \underline{254}$ & $\underline{0.99} \pm \underline{0.10}$ & $\underline{2.0K} \pm \underline{276}$ \\ \hline
SIASM             & $\bm{0.99} \pm \bm{0.02}$ & $\bm{2.4K} \pm \bm{147}$ & $\bm{0.99} \pm \bm{0.03}$ & $\bm{2.4K} \pm \bm{181}$ \\ \hline
\multirow{2}{*}{Methods} & \multicolumn{2}{c|}{$p=0.25$} & \multicolumn{2}{c|}{average performance} \\ \cline{2-5}
                  & \% $\uparrow$ & Steps $\uparrow$ & \% $\uparrow$ & Steps $\uparrow$ \\ \hline
ACRON-H           & $0.53 \pm \underline{0.07}$ & $1.2K \pm 1K$ & $0.37 \pm 0.08$ & $1.2K \pm 1K$ \\ \hline
SIASM-H           & $0.88 \pm 0.20$ & $1.6K \pm 886$ & $0.58 \pm \underline{0.18}$ & $1.5K \pm 961$ \\ \hline
ACRON             & $\underline{0.99} \pm 0.10$ & $\underline{2.0K} \pm \underline{305}$ & $\underline{0.93} \pm 0.18$ & $\underline{2.0K} \pm \underline{278}$ \\ \hline
SIASM             & $\bm{0.99} \pm \bm{0.03}$ & $\bm{2.4K} \pm \bm{267}$ & $\bm{0.99} \pm \bm{0.03}$ & $\bm{2.4K} \pm \bm{198}$ \\ \hline
\end{tabular}}
\caption{Total survival timesteps v.s. network influence ratio after reaching $|F|=|V|$. Bold: the best performance, underline: the second performance.}
\label{fig:ss}
\end{table}

% \begin{table}[t]
% \centering
% \caption{Survival timesteps after reaching $|S|=|V|$. \textbf{Bold}: the best performance, \underline{underline}: the second performance.}
% \resizebox{1\linewidth}{!}{
% \begin{tabular}{|c|c|c|c|c|}
% \hline
% Timesteps & $p=0.25$ & $p=0.5$ & $p=0.75$ & Average \\ \hline
% ACRON-H & $1.2K \pm 1K$ & $1.2K \pm 1K$ & $1.2K \pm 1K$ & $1.2K \pm 1K$ \\ \hline
% SIASM-H & $1.8K \pm 996$ & $1.1K \pm 1K$ & $1.6K \pm 886$ & $1.5K \pm 961$ \\ \hline
% ACRON & $\underline{2.1K} \pm \underline{254}$ & $\underline{2.0K} \pm \underline{276}$ & $\underline{2.0K} \pm \underline{305}$ & $\underline{2.0K} \pm \underline{278}$ \\ \hline
% SIASM & $\bm{2.4K} \pm \bm{147}$ & $\bm{2.4K} \pm \bm{181}$ & $\bm{2.4K} \pm \bm{267}$ & $\bm{2.4K} \pm \bm{198}$ \\ \hline
% Avg.($\%$) $\uparrow$ & $14.29\%$ & $20.00\%$ & $20.00\%$ & $20.00\%$ \\ \hline
% Dev.($\%$) $\downarrow$ & $42.13\%$ & $34.42\%$ & $12.46\%$ & $28.78\%$ \\ \hline
% \end{tabular}}
% \label{fig:ss}
% \end{table}

\subsubsection{Heterogeneous Datasets.}
Under heterogeneous settings, we maintain multi-relational social activities to obtain more comprehensive user embeddings and model adversarial socialbot behaviors based on the principles of structural information, as depicted in Fig. \ref{fig:SIASM}. 
The SIASM and other baselines are trained using the default training parameters, $p$ and $T_{max}$, which are consistent with homogeneous training.
The training curves of these models are presented in Fig. \ref{fig:training}, wherein the SIASM exhibits the fastest convergence and achieves the best performance with an episode reward of $828.61$ and an episode length of $110.61$. 
By effectively leveraging heterogeneous information from the original social networks, the SIASM and its variant, SIASM-H, demonstrate the highest and second-highest performances in terms of network influence ($840.89$ and $831.83$, respectively) as well as sustainable stealthiness ($80.0$ and $75.39$, respectively) when tested on real-life graphs.
% To further demonstrate the network influence of SIASM, we compare its network influence ratio with all baselines in Fig. \ref{fig:homo_test}.
% It is evident that the SIASM stably achieves a greater network influence while interacting with fewer social users, regardless of propagation probabilities.
% And the results of the sustainable stealthiness are shown in Table.\ref{fig:ss}, where our framework survives for an average of $2.4K$ timesteps, performing much better than other methods.

\subsubsection{Ablation Studies.}
\begin{figure}[t]
    \centering
    \includegraphics[width=1\columnwidth]{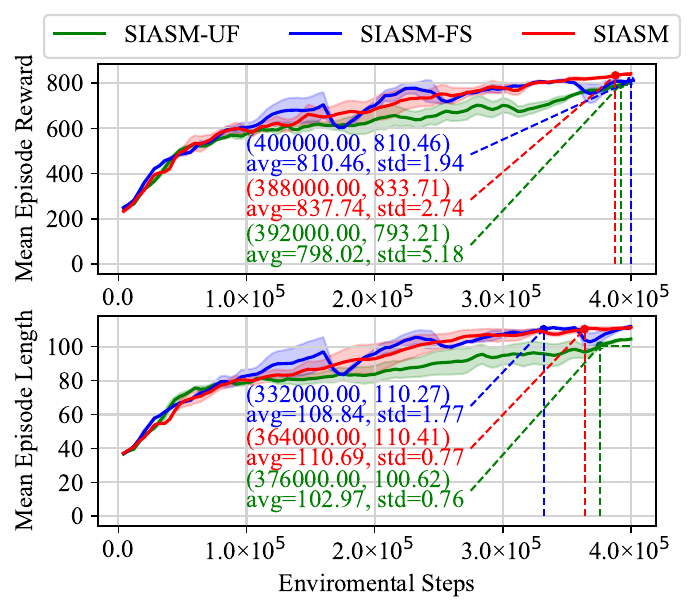}
    \caption{Learning curves of episode reward and length for ablation studies.}
    \label{fig:ablation}
\end{figure}

In this paper, we conduct ablation studies on homogeneous social networks to examine the effects of user filtration and follower selection stages in SIASM. 
The corresponding variants are referred to as SIASM-UF and SIASM-FS, respectively.
The results presented in Fig. \ref{fig:ablation} demonstrate that removing either the user filtration or follower selection stage decreases the overall quality and efficiency of policy learning.
These findings indicate that the filtration and selection stages based on the principles of structural information play a crucial role in enhancing the performance of adversarial modeling.

\section{Related Work} \label{section: related work}
\subsection{Adversarial Socialbot Modeling (ASM)}
Unlike traditional detection methods \cite{beskow2019its, yang2020scalable} that rely on static snapshots of features, ASM computationally models adversarial socialbot behaviors over time.
An evolution optimization algorithm is employed to generate various permutations from a pre-defined sequence of activities and select the best one to improve the accuracy of detectors \cite{cresci2019better}.
However, these permutations only provide static snapshots of behaviors and do not account for the socialbot's evolution.
To capture the temporal dynamics of socialbot, a general RL framework \cite{le2022socialbots} formulates adversarial behaviors as a Markov Decision Process (MDP).

\subsection{Structural Information Principles}
The concept of structural information was first introduced in 2016 by Li and Pan \cite{li2016structural}.
They proposed a metric that included definitions of structural entropy and partitioning tree, which can measure the dynamic complexity of networks and detect their natural hierarchical structure.
The one-dimensional structural entropy minimization principle was then used to identify subtypes of cancer cells by constructing cell sample networks \cite{li2016three}.
Later, Li et al. \cite{li2018decoding} decoded topologically associating domains of Hi-C data by minimizing high-dimensional structural entropy.
Recent advancements have widely applied the structural information principle across various domains, such as graph structure learning \cite{zou2023se}, skin lesion segmentation \cite{zeng2023unsupervised}, and dimensionality estimation for GNNs \cite{yang2023minimum}.
Our team defined state and action abstractions on the encoding trees to achieve efficient and effective general decision-making frameworks \cite{zeng2023effective, zeng2023hierarchical}.

\section{Conclusion} \label{section: conclusion}
This paper proposes a structural information principles-based framework SIASM for adversarial socialbots modeling to advance proactive detection.
To maximize network influence under robust detectors, an influence quantification method and a conditional structural entropy are designed to guide follower selection.
Evaluations of challenging tasks within homogeneous and heterogeneous social networks demonstrate that SIASM significantly improves network influence and sustainable stealthiness compared to state-of-the-art baselines.
In the future, we plan to extend our existing work by incorporating adversarial modeling of multiple socialbots and expanding proactive detection.

\section*{Acknowledgments}
The corresponding authors are Hao Peng and Angsheng Li. 
This work is supported by National Key R\&D Program of China through grant 2022YFB3104700, NSFC through grants 61932002 and 62322202, Beijing Natural Science Foundation through grant 4222030, and the Fundamental Research Funds for the Central Universities.

\bibliography{main}

\end{document}